\RequirePackage{lineno}
\documentclass[aps,twocolumn,superscriptaddress,showpacs,amsmath,amssymb,nofootinbib]{revtex4}
 \pdfoutput=1
 \usepackage{graphicx,epsf,amsmath}
 \usepackage[]{latexsym}
 \usepackage{amsmath}
 \usepackage{subcaption}
 \usepackage{float}
 \usepackage{color}
 
 \usepackage{dcolumn}   % needed for some tables
 \usepackage{bm}        % for math
 \usepackage{amssymb}   % for math
 \usepackage[utf8]{inputenc}

 \newcommand{\be}{\begin{eqnarray}}
 \newcommand{\ee}{\end{eqnarray}}

\usepackage{lineno,lipsum,everyhook}

\begin{document}
 \clearpage
 %\setpagewiselinenumbers
 %\modulolinenumbers[1]
 %\linenumbers

\title{UHECRs mass composition from $X_{\rm max}$ distributions}

 \author{Nicusor Arsene}
 \email[]{nicusorarsene@spacescience.ro}
 \affiliation{Institute of Space Science, P.O.Box MG-23, Ro 077125 
 Bucharest-Magurele, Romania}

 \author{Octavian Sima}
 \email[]{octavian.sima@partner.kit.edu}
 \affiliation{Physics Department, University of Bucharest, 
 Bucharest-Magurele, Romania}
 \affiliation{“Horia Hulubei” National Institute for Physics and Nuclear Engineering, Romania}
 \affiliation{Extreme Light Infrastructure - Nuclear Physics, ELI-NP, Ro 07725 Bucharest-Magurele, Romania}

 \date{\today}

\begin{abstract}
%\linenumbers
The atmospheric depth where the energy deposit profile of secondary particles from extensive air showers (EAS) reaches its maximum, $X_{\rm max}$, is related to the primary particle mass.
The mass composition of the ultra-high energy cosmic rays (UHECRs) can be inferred from measurements of $X_{\rm max}$ distributions in each energy interval, by fitting these distributions with Monte Carlo (MC) templates for four primary species (p, He, N and Fe). On the basis of simulations, we show that a high abundance of some intermediate elements in the $X_{\rm max}$ distributions, e.g. Ne or Si, may affect the quality of the fit and also the reconstructed fractions of different species with respect to their true values.
We propose a method for finding the "best combination" of elements in each energy interval from a larger set of primaries (p, He, C, N, O, Ne, Si and Fe) which best describes the $X_{\rm max}$ distributions. Applying this method to the $X_{\rm max}$ distributions measured by the Pierre Auger Observatory (2014), we found that the "best combination" of elements which best describe the data suggest the presence of Ne or Si in some low energy bins for the EPOS-LHC model.

\end{abstract}


\maketitle

\section{Introduction}
 %\setpagewiselinenumbers

The mass composition of the UHECRs is one of the most important ingredients needed when trying to elucidate the origin and acceleration mechanisms of these most energetic particles in the universe.
The most reliable observable from extensive air showers (EAS) used to infer the mass composition is the $X_{\rm max}$ parameter \cite{GH}, the atmospheric depth where the energy deposit profile of secondary particles reaches its maximum.
This parameter is related to the mass of the primary particle which initiate the shower, $\langle X_{\rm max}\rangle \propto - \ln A$, with larger mean values and dispersion for light primary particles in comparison with the heavier nuclei.
Experimentally, the mass composition of UHECRs was inferred from measurements of the first two moments of the $X_{\rm max}$ distributions ($\langle X_{\rm max}\rangle$ and $\sigma_{X_{\rm max}}$) as a function of the primary energy, by the Pierre Auger \cite{ThePierreAuger:2015rma, PhysRevLett.104.091101, Aab:2014kda}, High-Resolution Fly’s Eye (HiRes) \cite{PhysRevLett.104.161101} and the Telescope Array \cite{Abbasi:2018nun} Collaborations. Despite of the large data acquisition time (the Pierre Auger Observatory is operating since 2004) and large acceptance of the experiments, the reconstruction of the mass composition is  affected by large uncertainties mainly due to the unknown interaction cross sections at highest energies, experimental systematic uncertainties and poor statistics at the highest energies.    

In \cite{PhysRevD.90.122006} the Pierre Auger Collaboration show that using only the limited information given by the first two moments of the $X_{\rm max}$ distributions, degeneracies may be induced when interpreting the mass composition of a given $X_{\rm max}$ distribution, e.g. different mixes of primary particles can have identical mean and dispersion. 
To get information on fractions of individual nuclei, the Pierre Auger Collaboration used the entire shape of $X_{\rm max}$ distributions fitting them with MC templates for (p, He, N, Fe). The fits were performed with a binned maximum-likelihood method and the goodness of the fits was characterised with \textit{p-value}. 
With the use of this method the Auger
data for $E > 10^{17.8}$ eV could be described well with mixed compositions consisting of p, He and N (as representative for the intermediate mass elements), while fractions of Fe were close to zero in most of the energy bins.

In this work we show that fitting the $X_{\rm max}$ distributions with (p, He, N, Fe) elements, the fit quality is affected if some intermediate elements, e.g. Ne/Si are in fact present.
For that, we propose a method for finding the best combination from a larger set of possible elements (p, He, C, N, O, Ne, Si and Fe) to fit the data. Applying this method to the Auger data reported in \cite{Aab:2014kda} we observe a slight improvement of \textit{p-values} when Ne/Si are considered as additional fitting elements in some energy bins, especially at energies below the \textit{ankle} ($E < 10^{18.6}$ eV) where the statistics in the data is larger.

In Section \ref{sim} we describe the simulation procedure to obtain the $X_{\rm max}$ Probability Density Functions (PDFs) for each primary species in the energy range $\lg (E/\rm eV) = [17.8 - 19.3]$. In Section \ref{Ne-Si} we show the influence of Ne/Si abundance on the goodness of fit parameter \textit{p-value}. In Section \ref{fit} we present the fit results on $X_{\rm max}$ distributions measured by the Pierre Auger Observatory (until 2014), considering the elements which best describe the data. Section \ref{conclusions} concludes the paper.

%%--------------------------------------------------------------------------------

\section{Simulations}
\label{sim}

\begin{figure}
 \centering
\includegraphics[height=.27\textheight]{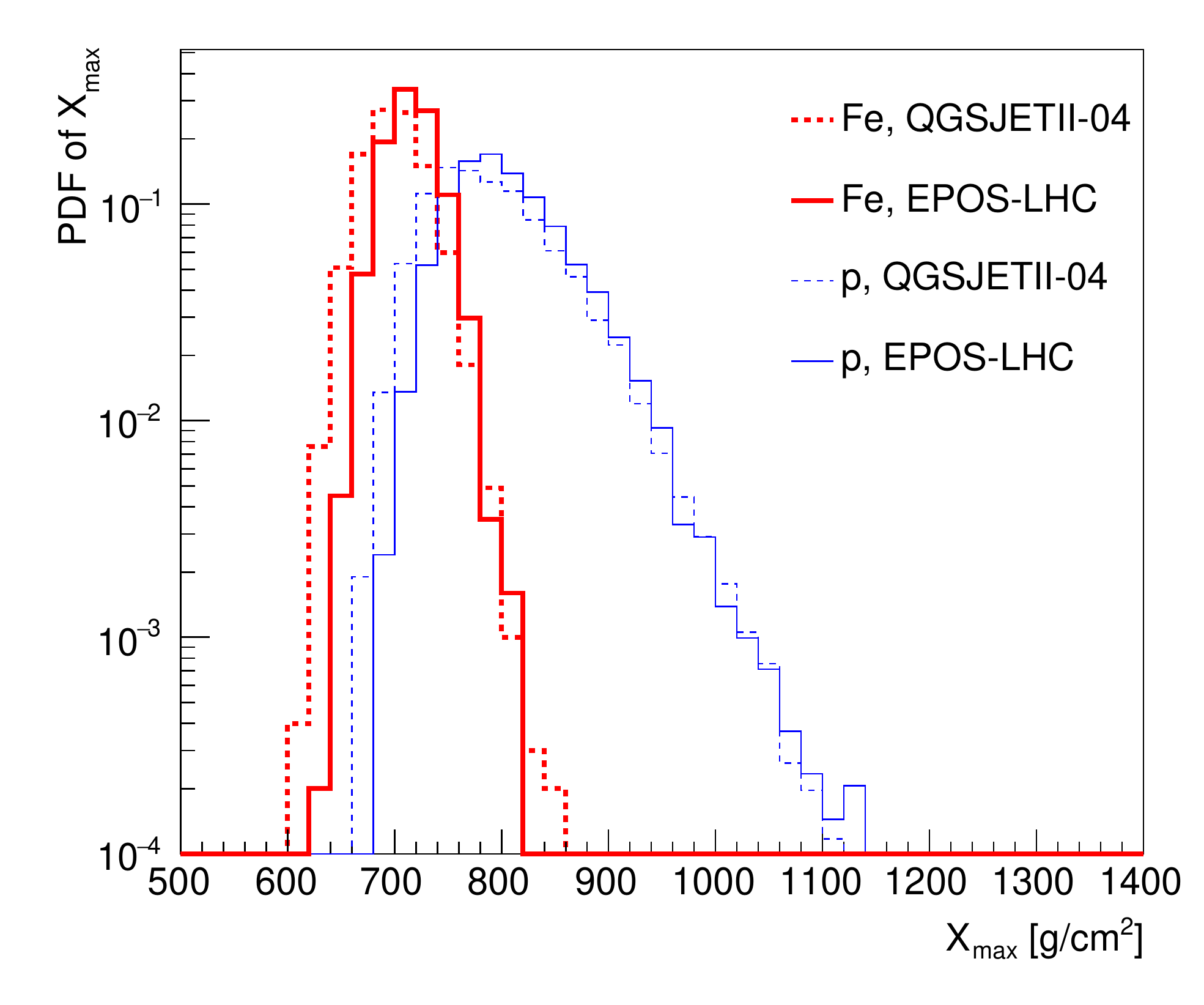}
\caption{PDFs of $X_{\rm max}$ for proton and iron induced showers considering EPOS-LHC and QGSJETII-04. The $X_{\rm max}$ values are obtained from CONEX simulations taking into account the experimental acceptance and resolution effects (Eq. (7) and (8) from \cite{Aab:2014kda}). }
\label{pdf}
\end{figure}

We used the CONEX v4r37 simulation code \cite{Pierog:2004re, Bergmann:2006yz} to generate the $X_{\rm max}$ distributions for each element (p, He, C, N, O, Ne, Si and Fe) in 15 energy intervals of 0.1 in $\log(E/eV)$ starting from $E = 10^{17.8}$ eV up to $E = 10^{19.3}$ eV. Three high energy hadronic interaction models were employed, EPOS-LHC \cite{PhysRevC.74.044902}, QGSJETII-04 \cite{PhysRevD.74.014026} and Sibyll 2.1 \cite{PhysRevD.80.094003}. The zenith angle of the showers were sampled from an isotropic distribution in the interval $\theta = [0^{\circ} - 60^{\circ}]$.
The statistics of the simulation data set consists in $10^4 - 10^{5}$ events per each primary species per hadronic interaction model in each energy interval.
A PDF of $X_{\rm max}$ for a nuclear species in a given energy interval consists in a binned $X_{\rm max}$ distribution in the range $[0 - 2000]$ g/cm$^{2}$ with a bin width $= 20$ g/cm$^{2}$. 
The true $X_{\rm max}$ values given by the CONEX simulations were modified to account for the detector acceptance and experimental resolution (Eq. (7) and (8) from \cite{Aab:2014kda}).
An example of PDFs of $X_{\rm max}$ for proton and iron induced showers in the energy interval $\lg (E/ \rm eV) = [19.0 - 19.1]$ for two hadronic interaction models is presented in Figure\ref{pdf}.

We will use these PDFs in the next section to generate random $X_{\rm max}$ distributions with different mixes of primary particles to observe the behavior of the goodness of fit estimator \textit{p-value} as a function of different prior abundances, when the $X_{\rm max}$ distributions are fitted with the four fixed PDFs (p, He, N and Fe).
%%--------------------------------------------------------------------------------

\section{Influence of N\lowercase{e}/S\lowercase{i} on the goodness of fit}
\label{Ne-Si}

The results on mass composition of primary cosmic rays at energies $E > 10^{17.8}$ eV reported in \cite{PhysRevD.90.122006} indicate a modulation of the abundances of primary protons, He and N nuclei as a function of energy. The experimental $X_{\rm max}$ distributions in each energy interval were fitted with four primary PDFs (p, He, N and Fe) following a binned maximum-likelihood procedure.
Different astrophysical models suggest a variation of the abundance of different elements as a function of energy below and above the \textit{ankle} \cite{PhysRevD.72.081301, PhysRevD.74.043005, 2012APh....39..129A, PhysRevD.92.123001}. In such a scenario, the observed modulation of the reconstructed fractions might be biased as a consequence of a high abundance of an intermediate element not included into the fitting procedure, in the case when the $X_{\rm max}$ distributions are fitted with the same fixed four species (p, He, N and Fe) over the entire energy range.

We performed the following test. Using individual $X_{\rm max}$ values obtained from simulations as explained in Section \ref{sim}, we build $X_{\rm max}$ distributions for each energy bin considering random abundances of 8 primary species (p, He, C, N, O, Ne, Si and Fe). We generated a large number of such distributions ($3 \times 10^{4}$) to ensure that we cover all possible mixes. The statistics in each distribution is of the same magnitude as in the Auger data. Then, using a binned maximum-likelihood procedure we fit these $X_{\rm max}$ distributions with 4 PDFs (p, He, x, Fe), where x was varied from C to Si.

The minimizing quantity, $-\ln L$, in this fitting procedure is defined as:
\begin{equation}
\label{logl}
 -\ln L = \sum_{i} y_i - n_i + n_i \ln(n_i / y_i) ,
\end{equation}
where $n_i$ stands for the measured counts in the "$i$"-th bin of an $X_{max}$ distribution and $y_i$ represents the MC prediction for that bin \cite{Baker:1983tu}. 
The \textit{p-value} parameter represents the probability of obtaining a worse fit than that observed, even if the distribution predicted by the fit results is correct:
\begin{equation}
\label{pval}
\textit{p-value} = 1 - \Gamma\left(\frac{ndf}{2}, \frac{\chi^2}{2}\right), 
\end{equation}
where $\Gamma$ is the incomplete gamma function, $ndf$ represents the number of degrees of freedom, and $\chi^2$ represents the sum of the square of residuals using the parameters computed by the likelihood method. Note that the \textit{p-values} calculated using Eq. \ref{pval} differ from those calculated in \cite{PhysRevD.90.122006}. We make the approximation that $L$ behaves like a $\chi^2$ variable while in \cite{PhysRevD.90.122006} the 
\textit{p-value} parameters are calculated in a more realistic way, using mock data sets of the predicted fractions with size equal to the real data sets. Even if the absolute \textit{p-values} might be affected by this approximation we consider that
the relative variation of \textit{p-values} with the components included in the fit is significant. In addition, we mention that the \textit{p-values} obtained by us with Eq. \ref{pval} do not differ significantly from those obtained with the
method used in \cite{PhysRevD.90.122006}, therefore we consider that the main conclusion of this paper will be not affected by this choice.

\begin{figure}
 \centering
\includegraphics[height=.32\textheight]{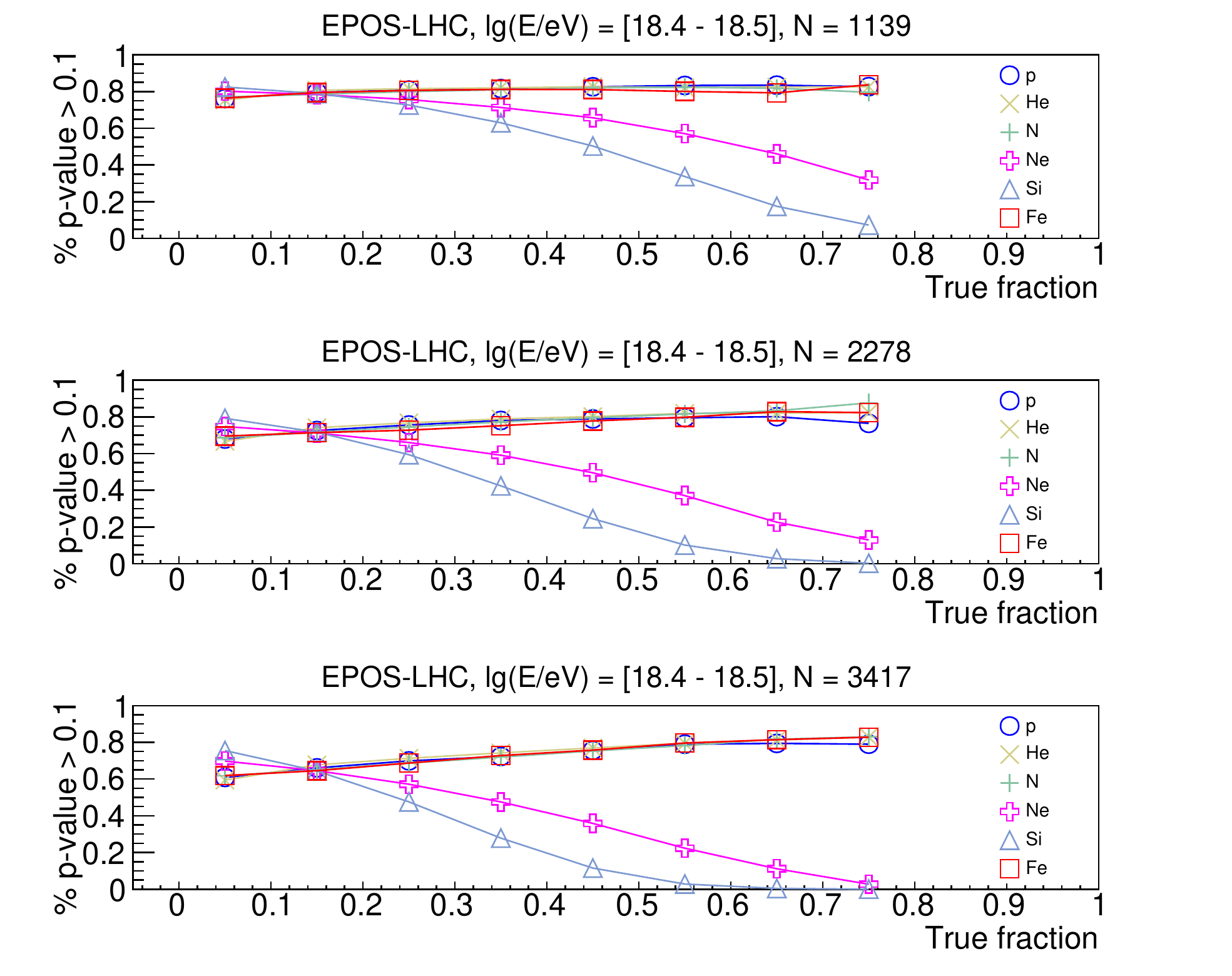}
\caption{Fraction of events with a \textit{p-value} greater than 0.1 as a function of prior abundance of different species. The $X_{\rm max}$ distributions correspond to the energy interval $\lg (E/\rm eV) = [18.4 - 18.5]$ considering EPOS-LHC. The fitting function includes 4 elements (p, He, N and Fe). The statistics in the $X_{\rm max}$ distributions is $N = N_{Auger} = 1139$ \textit{(top)}, $N = 2 N_{Auger} = 2278$ \textit{(middle)} and $N = 3 N_{Auger} = 3417$ \textit{(bottom)}.}
\label{pval-4}
\end{figure}

\begin{figure}
 \centering
\includegraphics[height=.32\textheight]{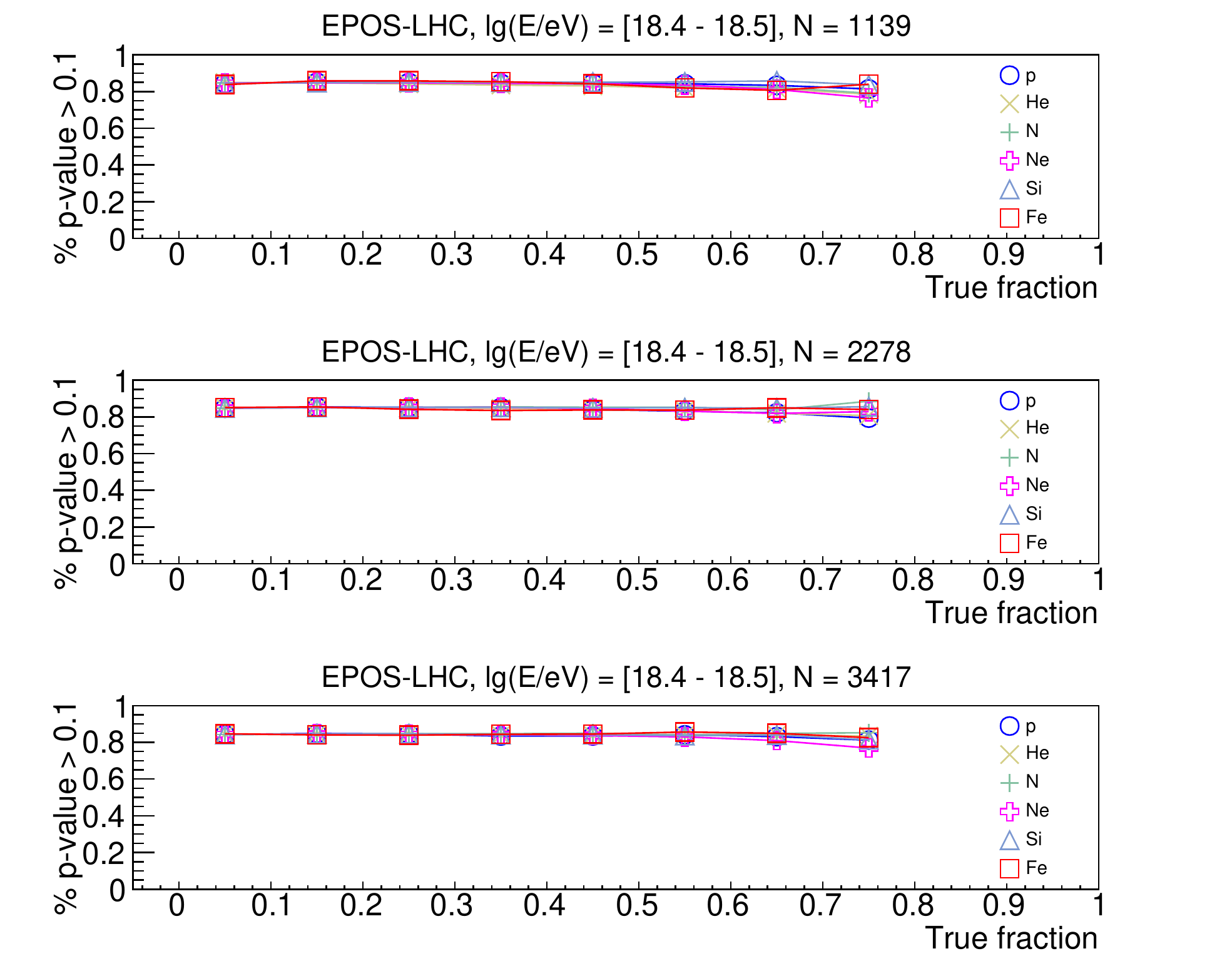}
\caption{Fraction of events with a \textit{p-value} greater than 0.1 as a function of prior abundance of different species. The $X_{\rm max}$ distributions correspond to the energy interval $\lg (E/\rm eV) = [18.4 - 18.5]$ considering EPOS-LHC. The fitting function includes 5 elements (p, He, N, Si and Fe). The statistics in the $X_{\rm max}$ distributions is $N = N_{Auger} = 1139$ \textit{(top)}, $N = 2 N_{Auger} = 2278$ \textit{(middle)} and $N = 3 N_{Auger} = 3417$ \textit{(bottom)}.}
\label{pval-5}
\end{figure}

Indeed, the best results were obtained when the distributions were fitted with p, He, Fe and any of CNO nuclei. Further, we tried to check what is the capability of this fitting method to reconstruct these 4 abundances (p, He, N, Fe) if one of the primary species has a high prior abundance.
An example of the evolution of the fit quality as a function of different abundance of nuclear species in the $X_{\rm max}$ distributions is represented in Figure \ref{pval-4} for the energy interval $\lg (E/\rm eV) = [18.4 - 18.5]$ for EPOS-LHC. We considered the actual statistics measured by Auger in this energy interval, N = 1139 \textit{(upper panel)} and the case in which we double and triple the number of events in distributions \textit{(middle panel} and \textit{bottom panel} respectively). The results from Figure \ref{pval-4} and \ref{pval-5} can be interpreted as follows: the first blue circle stands for the case in which the true fraction of protons in $X_{\rm max}$ distributions was in the interval $[0 - 0.1]$ while the rest 7 elements had random abundances. 
The second blue circle stands for the case when the true fraction of protons was in the range $[0.1 - 0.2]$ and so on. Similarly, the green "x" cross symbol stands for the case in which the true fraction of He in $X_{\rm max}$ distributions was in the interval $[0 - 0.1]$ while the rest 7 elements had random abundances and so on.
It was convenient to quantify the quality of fit as fraction of events with \textit{p-value}$ > 0.1$. 

As we can see in Figure \ref{pval-4}, the probability of obtaining a good \textit{p-value} decreases with the increase of abundances of Ne or Si and with increase of statistics in $X_{\rm max}$ distributions, when the fitting procedure includes only four PDFs (p, He, N and Fe). Moreover, we found that the reconstructed fractions of protons, He and Fe differ from the true fractions by up to $20 \%$ in some cases (i.e. when the abundance of Ne or Si is $> 40 \%$). 
When fitting the same distributions with 5 elements including Si (p, He, N, Si and Fe), the fit quality is not affected by the higher prior abundances of Ne or Si. These results are presented in Figure \ref{pval-5}.  We observed that in the energy bins where the statistics is very small (e.g. $\lg (E/\rm eV) = [19.2 - 19.3]$, $N = 87$), the higher abundance of Ne or Si does not affect the quality of the fit. For these energies the
reliable estimations on the mass composition can not be obtained due to poor statistics available in data.

%%--------------------------------------------------------------------------------

\section{Fitting Auger $X_{\rm max}$ distributions}
\label{fit}

We fit the experimental $X_{\rm max}$ distributions measured at the Pierre Auger Observatory \cite{Aab:2014kda}, with the four fixed PDFs (p, He, N and Fe) on the entire energy range $\lg (E/\rm eV) = [17.8 - 19.3]$. 
The results we have obtained are in a very good agreement with those reported in \cite{PhysRevD.90.122006}, since we build our PDFs based on the same version of CONEX code, employing the same versions of hadronic interaction models and considering the same binned maximum-likelihood fitting procedure. 

We found that the most appropriate approach to fit the experimental $X_{\rm max}$ distributions in each energy interval is to consider all possible combinations of PDFs from a larger set of nuclear elements (p, He, C, N, O, Ne, Si and Fe) and then to find the "best combination" of elements which best describe the data. 
Thus, the number of elements from a "best combination" may vary between 1 and 8.
We will refer from now on to this fitting approach as "best combination". 

It is worth mentioning that in the minimization procedure of the log-likelihood (Eq. \ref{logl}) we do not neglect the empty bins and the $ndf$ parameter is calculated as the number of bins in the $X_{max}$ distribution minus the number of parameters considered in the fit. Therefore, the computation of the \textit{p-value} parameter takes into account the number of parameters considered in the fit.

\begin{figure}
 \centering
\includegraphics[height=.175\textheight]{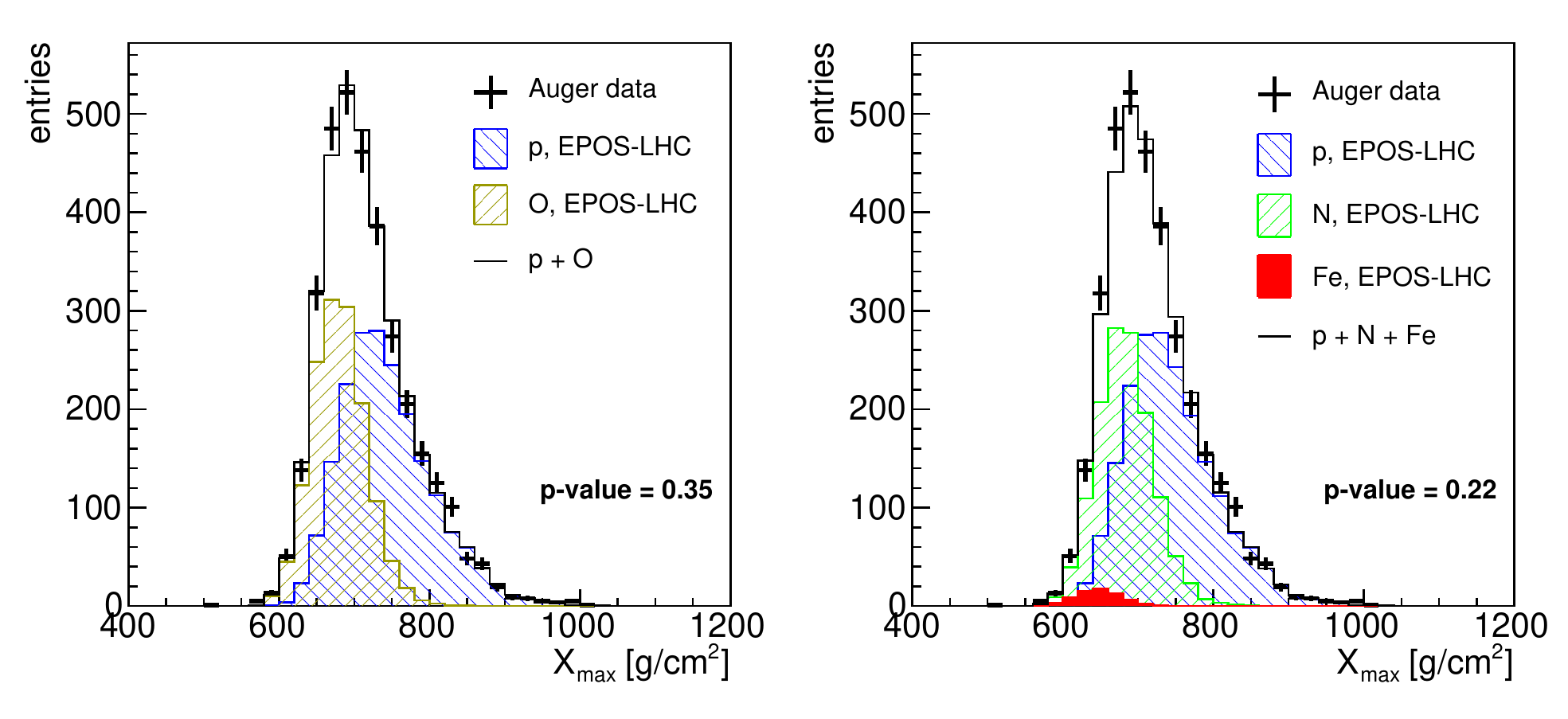}
\caption{$X_{\rm max}$ distribution measured at the Pierre Auger Observatory in the energy interval $\lg (E/\rm eV) = [17.9 - 18.0]$. The reconstructed fractions predicted by the "best combination" fitting procedure are displayed on the \textit{(left)} with \textit{p-value}$ = 0.35$. The reconstructed fractions obtained by fitting the same distribution with 4 PDFs (p, He, N and Fe) are displayed on the \textit{(right)} (\textit{p-value}$= 0.22$). All the PDFs are obtained considering the EPOS-LHC interaction model.}
\label{fits_17_9}
\end{figure}

In Figure \ref{fits_17_9} we give an example of a $X_{\rm max}$ distribution measured by the Pierre Auger Observatory in the energy interval $\lg (E/\rm eV) = [17.9 - 18.0]$. We found that the "best combination" (Figure \ref{fits_17_9} \textit{left}) suggests that the shape of the distribution is best described only by two elements, protons and O, with \textit{p-value}$= 0.35$, for the case of EPOS-LHC model. In  Figure \ref{fits_17_9} \textit{right} we present the results obtained by fitting the same $X_{\rm max}$ distribution with 4 PDFs (p, He, N and Fe). In this case we obtain a worse \textit{p-value}$= 0.22$.

A direct comparison of the two fitting procedures is presented in Figure \ref{epos}, \ref{qgsj} and \ref{sibyll} for EPOS-LHC, QGSJETII-04 and Sibyll 2.1 for the entire energy range. In the case of QGSJETII-04 (Figure \ref{qgsj}) and Sibyll 2.1 (Figure \ref{sibyll}) we observe negligible differences if the $X_{\rm max}$ distributions are fitted with the four fixed PDFs (p, He, N and Fe) or if we use the "best combination" of elements. 
The only modification consists in a slight improvement of the \textit{p-value} parameter over the entire energy range for the "best combination" case. 
Important to mention that the number of elements from the "best combination" consists in 2 or 3 elements over the entire energy range for each hadronic interaction model.
The error bars (statistical uncertainties) of the fitted fractions should not be compared with those from \cite{PhysRevD.90.122006} since they are computed considering different methods. We have employed the MINOS technique based on $\Delta L = 1/2$ rule, while in \cite{PhysRevD.90.122006} the authors used the Feldman-Cousins procedure in which the parameter uncertainties are computed in a more rigorous way by enforcing unitarity. Most likely the uncertainties from Figures \ref{epos}, \ref{qgsj} and \ref{sibyll} from our manuscript are underestimated.

The most interesting aspect is observed in the case of EPOS-LHC model (Figure \ref{epos}) at the lower energies. We found that for some energy intervals, e.g. $\lg (E/\rm eV) = [18.1 - 18.2]$, $[18.4 - 18.5]$,  the "best combination" suggest the presence of Ne or Si in Auger data with a slight improvement of the \textit{p-value} parameter.
This aspect is in agreement with our results from Section \ref{Ne-Si}, where we found that a high prior abundance of Ne or Si ($ > 20 \% $) may affect the quality of fit if the $X_{\rm max}$ distributions are fitted with the four PDFs (p, He, N and Fe). 
Without making speculations, one can consider that the results presented in this paper could be a hint for the presence of the heavier elements ($20 < A < 39$) around the \textit{ankle}, as predicted in \cite{PhysRevD.92.123001}.

\begin{widetext}

\begin{figure}
 \centering
\includegraphics[height=.8\textheight]{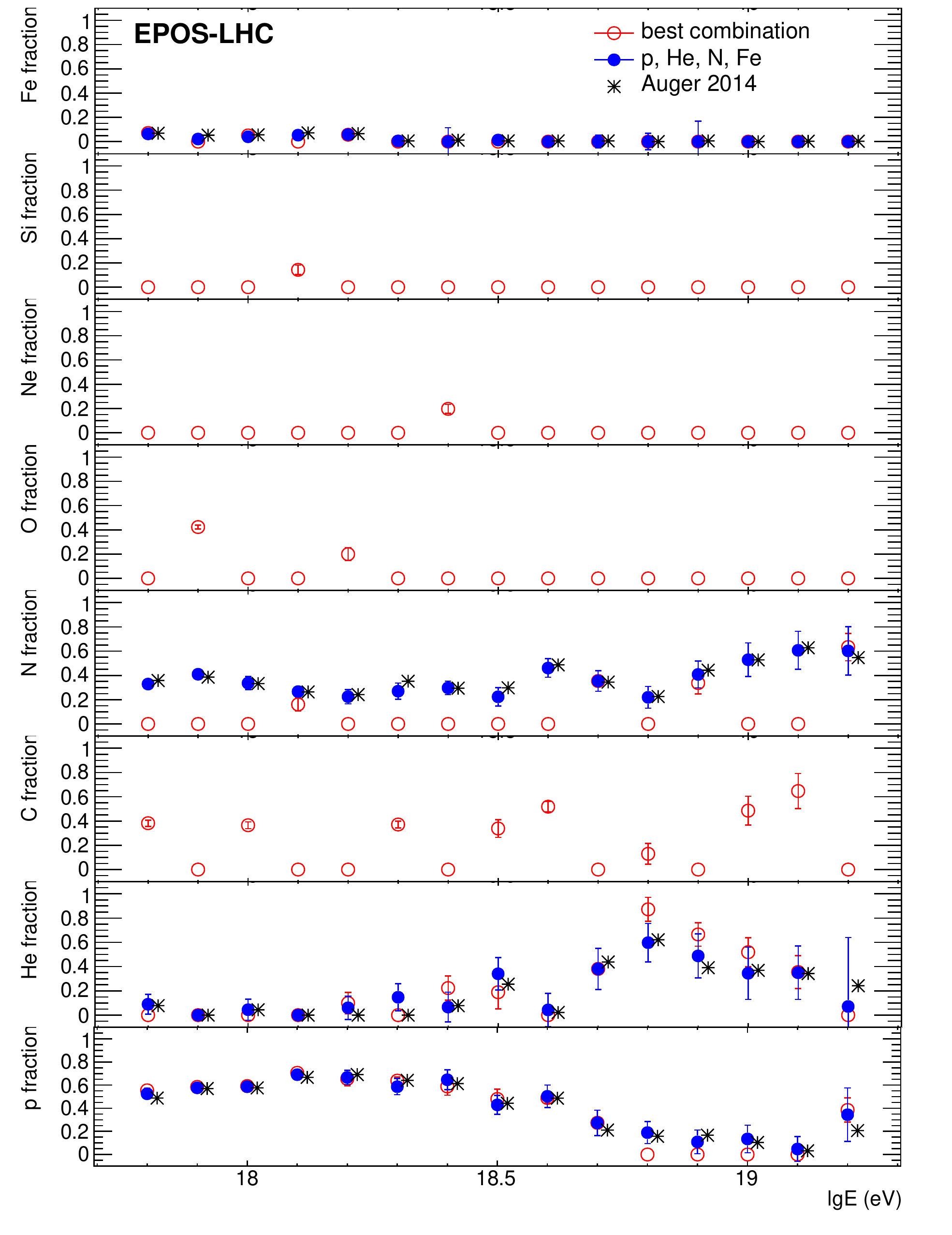}
\caption{Fitted fractions in each energy interval considering EPOS-LHC model. Blue full circles stand for the fitting method which uses only four fixed species (p, He, N and Fe) on the entire energy range. Red circles represent the fitted fractions found for the "best combination" method and black stars stand for Auger 2014 results \cite{PhysRevD.90.122006}. }
\label{epos}
\end{figure}

\begin{figure}
 \centering
\includegraphics[height=.8\textheight]{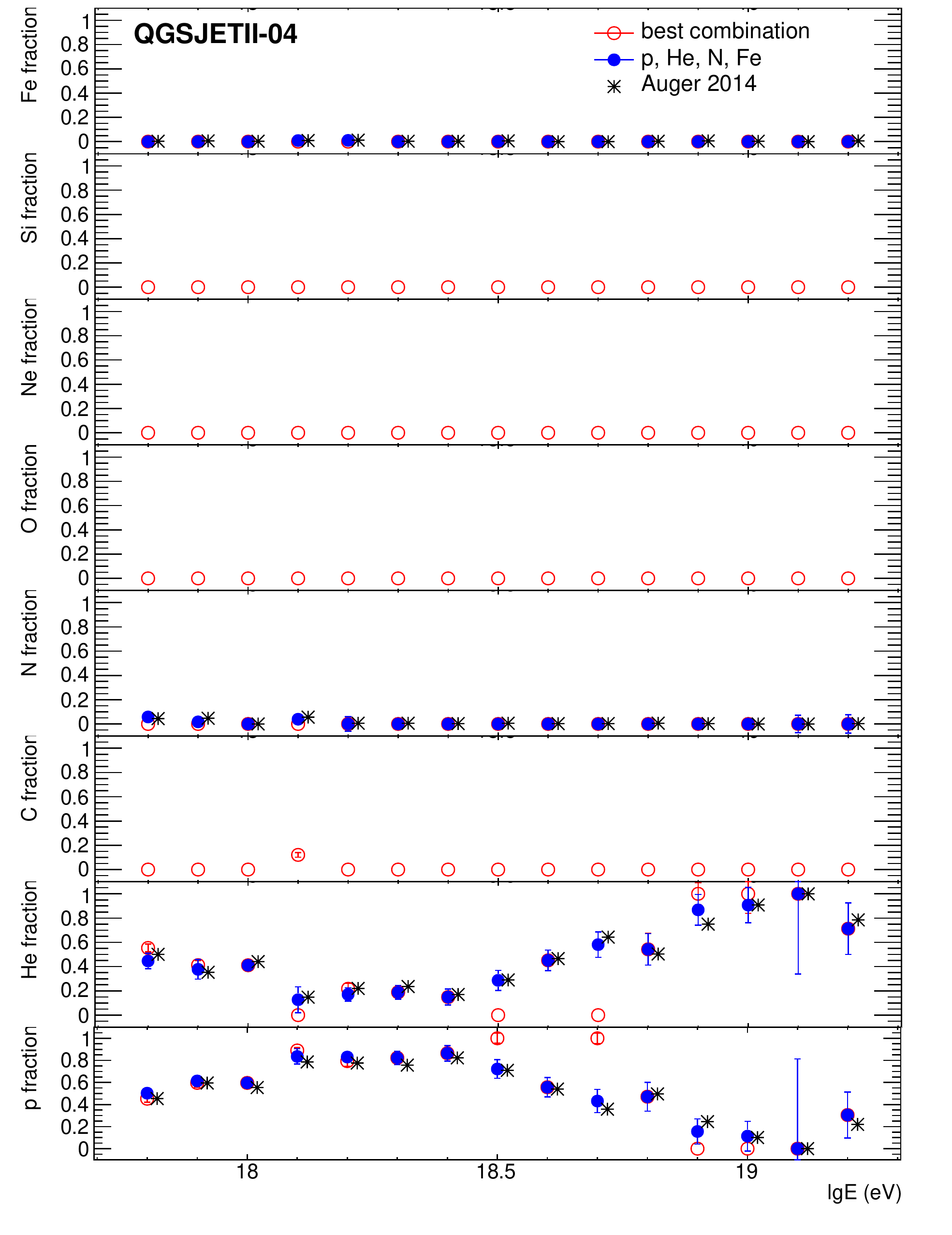}
\caption{Fitted fractions in each energy interval considering QGSJETII-04 model. Blue full circles stand for the fitting method which uses only four fixed species (p, He, N and Fe) on the entire energy range. Red circles represent the fitted fractions found for the "best combination" method and black stars stand for Auger 2014 results \cite{PhysRevD.90.122006}. }
\label{qgsj}
\end{figure}

\begin{figure}
 \centering
\includegraphics[height=.8\textheight]{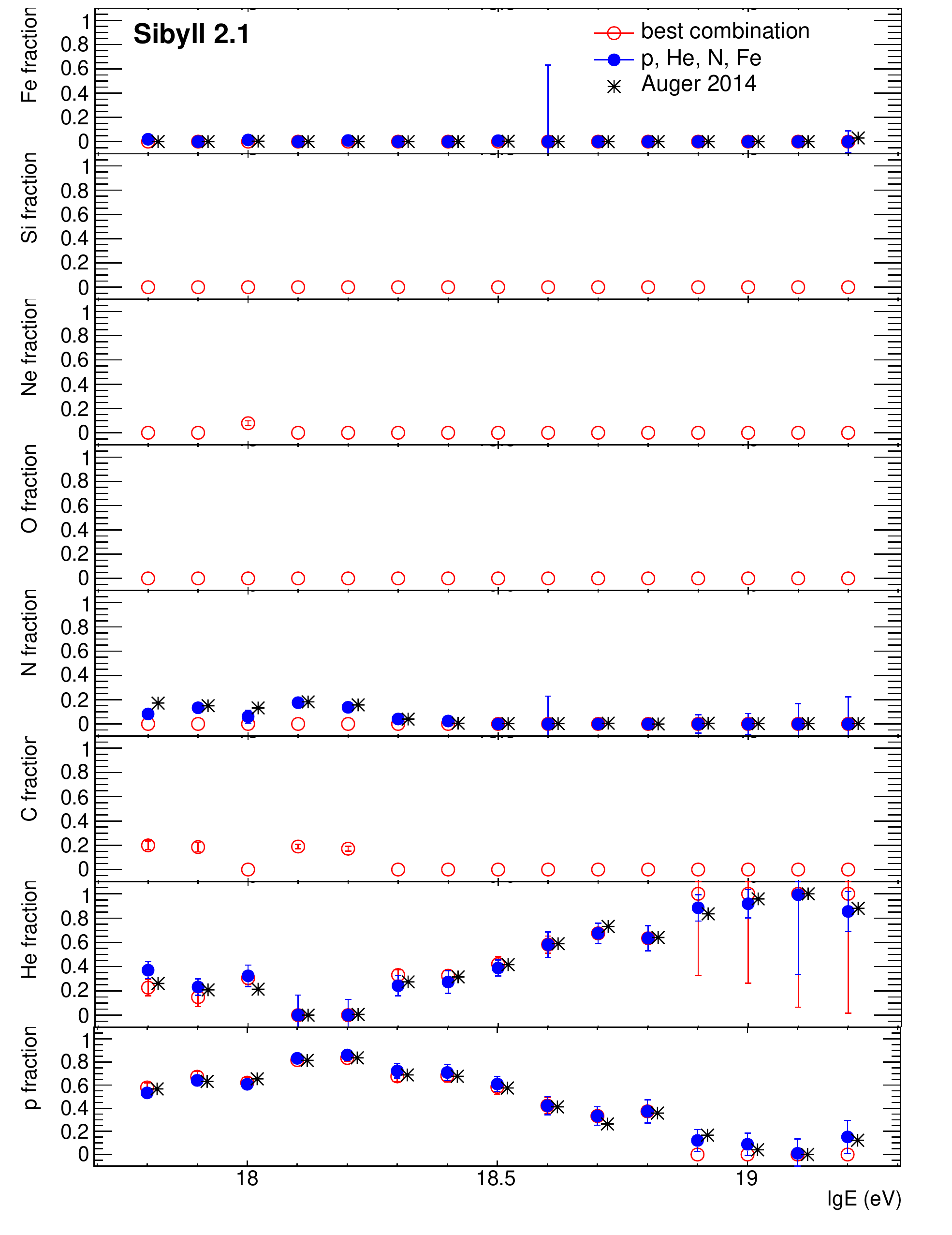}
\caption{Fitted fractions in each energy interval considering Sibyll 2.1 model. Blue full circles stand for the fitting method which uses only four fixed species (p, He, N and Fe) on the entire energy range. Red circles represent the fitted fractions found for the "best combination" method and black stars stand for Auger 2014 results \cite{PhysRevD.90.122006}.}
\label{sibyll}
\end{figure}

\end{widetext}

%%--------------------------------------------------------------------------------

\section{Discussions and conclusions}
\label{conclusions}

In this paper we investigated the capability to infer the mass composition of the primary UHECRs from measurements of $X_{\rm max}$ distributions. Using simulated $X_{\rm max}$ distributions for a large set of primary species (p, He, C, N, O, Ne, Si and Fe), we build $X_{\rm max}$ distributions with random mixes of elements for each energy interval in the energy range $\lg (E/\rm eV) = [17.8 - 19.3]$. We found that a high prior abundance of Ne or Si can bias the reconstructed fractions of elements if the distributions are fitted with four fixed PDFs (p, He, N and Fe). 
We found that the fit quality decreases with increasing the Ne/Si abundance and with increasing the statistics in the $X_{\rm max}$ distributions.

We proposed an alternative approach to infer the mass composition from the $X_{\rm max}$ distributions which finds the "best combination" of elements best describing the distributions from a larger set of primaries. 
Applying this method to the $X_{\rm max}$ distributions measured by the Pierre Auger Observatory until 2014, it was shown that in some low energy bins, only for the EPOS-LHC model the "best combination" of elements suggests the presence of Ne or Si, with a slight improvement of the \textit{p-value} parameter. 
Since we have shown using simulations that a high Ne/Si prior abundance will affect the fit quality if the $X_{\rm max}$ distribution is fitted with four PDFs (p, He, N and Fe), we consider that it is important to take into account further elements in future studies. 

%%--------------------------------------------------------------------------------

\subsection*{Acknowledgments}

We would like to thank our colleagues from the Pierre Auger Collaboration for many interesting and useful discussions. Special thanks to Alexey Yushkov for many important suggestions which helped us to improve the paper. 
N. A. acknowledges financial support from the LAPLAS VI program of the Romanian National Authority for Scientific Research (CNCS-UEFISCDI). The work of O. S. was supported by a grant of the Romanian Ministery of Research and Innovation, CCCDI - UEFISCDI, project number PN-III-P1-1.2 PCCDI-2017-0839/19PCCDI/2018, within PNCDI III.

\bibliography{Xmax_fit}

\end{document}